%% file: main.tex
\documentclass[sigconf]{acmart}
\acmConference[EASE 2025]{The 29th International Conference on Evaluation and Assessment in Software Engineering}{17–20 June, 2025}{Istanbul, Turkey}

\input{preamble.tex}

\begin{document}

\title{Evaluating Mutation-based Fault Localization\\ for Quantum Programs}

\input{author_info.tex}

\begin{abstract}
\input{section/abstract.tex}
\end{abstract}




\maketitle


\input{section/introduction.tex}
\input{section/background.tex}
\input{section/study_design.tex}
\input{section/results.tex}
\input{section/discussion.tex}
\input{section/validity.tex}
\input{section/conclusion.tex}

\begin{acks}
  We gratefully acknowledge the financial support of:
  (1) \grantsponsor{JST}{Japan Science and Technology Agency (JST)}{https://www.jst.go.jp/EN/} as part of Adopting Sustainable Partnerships for Innovative Research Ecosystem (ASPIRE), Grant Number \grantnum{JST}{JPMJAP2415};
  (2) \grantsponsor{Inamori}{the Inamori Research Institute for Science}{https://www.inamori-f.or.jp/en/inaris} for supporting Yasutaka Kamei via the \grantnum{Inamori}{InaRIS Fellowship}.
\end{acks}
  
\bibliographystyle{ACM-Reference-Format}
\bibliography{reference}

\end{document}

%% file: preamble.tex
\AtBeginDocument{%
  \providecommand\BibTeX{{%
    \normalfont B\kern-0.5em{\scshape i\kern-0.25em b}\kern-0.8em\TeX}}}

\usepackage{float}
\usepackage{hyperref}
\usepackage{booktabs}
\usepackage{algorithmic}
\usepackage{textcomp}
\usepackage{xcolor}
\usepackage{soul}
\usepackage{url}
\usepackage{xspace}
\usepackage{multirow}
\usepackage{enumitem}
\usepackage{colortbl}
\usepackage[skins,breakable]{tcolorbox}
\usepackage{framed}
\usepackage{hhline}
\usepackage{threeparttable}
\usepackage{url}
\usepackage{subcaption}
\usepackage{comment}
\usepackage{wrapfig}
\usepackage{subcaption}
\usepackage{caption}
\usepackage{xparse}
\usepackage{listings}
\tcbuselibrary{raster,skins}

\def\fig#1{Figure \ref{#1}}
\def\tab#1{Table \ref{#1}}
\def\sec#1{Section~\ref{#1}}
\def\et{et\ al.\xspace}
\def\eg{e.g.\xspace}
\def\ie{i.e.\xspace}

\definecolor{mygray}{gray}{0.9}
\definecolor{ForestGreen}{HTML}{288c66}
\definecolor{MyBlue}{HTML}{00008b}
\setulcolor{MyBlue}

\newcounter{examplecounter}[subsection]
\renewcommand{\theexamplecounter}{\arabic{section}.\arabic{subsection}.\arabic{examplecounter}}
\newlist{stepitemize}{itemize}{1}
\setlist[stepitemize,1]{leftmargin=1.45cm}

\newlist{phaseitemize}{itemize}{1}
\setlist[phaseitemize,1]{leftmargin=1.70cm}

\NewDocumentCommand{\todo}{o}{
  \fcolorbox{red}{gray!25}{\small\textcolor{red}{\tt{TODO\IfNoValueTF{#1}{}{: #1}}}}
}

\def\summarybox#1#2{
\begin{oframed}
\noindent \textbf{#1:}
#2
\end{oframed}
}

\newcommand{\tm}{\textregistered~}
\newcommand{\tms}{\textregistered}

\definecolor{keyword}{HTML}{F58220}  
\definecolor{comment}{rgb}{0.5, 0.5, 0.5}    
\definecolor{string}{HTML}{EE7800}     
\definecolor{identifier}{HTML}{583F99} 

\newif\ifhighlight
\highlightfalse 

\newcommand{\revision}[1]{\ifhighlight\textcolor{red}{#1}\else#1\fi}

%% file: author_info.tex
\author{Yuta Ishimoto}
\authornotemark[1]
\orcid{0009-0006-2606-5701}
\affiliation{%
\institution{Kyushu University}
\state{Fukuoka}
\country{Japan}
}
\email{ishimoto@posl.ait.kyushu-u.ac.jp}

\author{Masanari Kondo}
\orcid{0000-0002-6317-7001}
\affiliation{%
  \institution{Kyushu University}
  \state{Fukuoka}
  \country{Japan}
}
\email{kondo@ait.kyushu-u.ac.jp}

\author{Naoyasu Ubayashi}
\orcid{0000-0003-1578-7195}
\affiliation{%
  \institution{Waseda University}
  \state{Tokyo}
  \country{Japan}
}
\email{ubayashi@acm.org}

\author{Yasutaka Kamei}
\orcid{0000-0002-7058-1045}
\affiliation{%
  \institution{Kyushu University}
  \state{Fukuoka}
  \country{Japan}
}
\email{kamei@ait.kyushu-u.ac.jp}

\author{Ryota Katsube}
\affiliation{%
  \institution{Research \& Development Group, Hitachi, Ltd.}
  \state{Kanagawa}
  \country{Japan}
}
\email{ryota.katsube.tt@hitachi.com}

\author{Naoto Sato}
\affiliation{%
  \institution{Research \& Development Group, Hitachi, Ltd.}
  \state{Kanagawa}
  \country{Japan}
}
\email{naoto.sato.je@hitachi.com}

\author{Hideto Ogawa}
\affiliation{%
  \institution{Research \& Development Group, Hitachi, Ltd.}
  \streetaddress{744 Motooka}
  \state{Kanagawa}
  \country{Japan}
}
\email{hideto.ogawa.cp@hitachi.com}

\renewcommand{\shortauthors}{Ishimoto, et al.}

%% file: section/abstract.tex

Quantum computers leverage the principles of quantum mechanics to execute operations. 
They require \emph{quantum programs} that define operations on quantum bits (\emph{qubits}), the fundamental units of computation.
Unlike traditional software development, the process of creating and debugging quantum programs requires specialized knowledge of quantum computation, making the development process more challenging.

In this paper, we apply and evaluate \textit{mutation-based fault localization (MBFL)} for quantum programs with the aim of enhancing debugging efficiency.
We use \textit{quantum mutation operations}, which are specifically designed for quantum programs, to identify faults.
Our evaluation involves 23 real-world faults and 305 artificially induced faults in quantum programs developed with Qiskit\tms.
The results show that real-world faults are more challenging for MBFL than artificial faults.
In fact, the median EXAM score\revision{, which represents the percentage of the code examined before locating the faulty statement (lower is better)}, is 1.2\% for artificial benchmark and 19.4\% for the real-world benchmark in the worst-case scenario.
Our study highlights the potential and limitations of MBFL for quantum programs, considering different fault types and mutation operation types.
Finally, we discuss future directions for improving MBFL in the context of quantum programming.

%% file: section/introduction.tex
\section{Introduction}\label{sec:introduction}
Quantum computers fundamentally differ from classical computers by leveraging the principles of quantum mechanics to perform computations~\cite{nielsen2010quantum}.
As quantum computing research advances, interest in \textit{quantum programs} has grown~\cite{ramalho2024arxiv}.
A quantum program comprises a sequence of operations on quantum bits (\emph{qubits}), where each operation is known as a \textit{quantum gate}.

The process of creating and debugging quantum programs is inherently challenging, as it requires specialized knowledge of quantum computing.
For example, qubits cannot be copied like classical bits due to the no-cloning theorem~\cite{nielsen2010quantum}.
Moreover, not all developers are necessarily well-versed in quantum computing~\cite{shaydulin2020icsew}.
Consequently, such quantum-specific aspects lead to issues such as code smells~\cite{chen2023icse} and technical debt~\cite{ishimoto2024apsec}.

In this study, we apply and evaluate \textit{mutation-based fault localization (MBFL)} for quantum programs with the aim of enhancing debugging efficiency.
Fault localization is a technique for identifying faults within a program~\cite{wong2016tse}.
An effective fault localization method can assist developers~\cite{kochhar2016issta}, especially those who are not experts in quantum computing~\cite{shaydulin2020icsew}.

MBFL leverages the results of mutation analysis for quantum programs.
Mutation analysis generates \textit{mutants}, systematically modified versions of a program.
In classical programs, this technique has been used not only for testing but also for fault localization~\cite{moon2014icst}.
While numerous studies have explored mutation testing for quantum programs~\cite{ali2021icst,mendiluze2021ase,fortunato2022tqe, wang2022gecco}, its potential for fault localization remains unclear.
These studies use \emph{quantum mutation operations} (\eg, deleting quantum gates), specifically designed for quantum programs. 
We believe that these operations hold promise for enhancing fault localization in quantum programs.

Our evaluation involves 23 real-world faults and 305 artificial faults in quantum programs developed with Qiskit\tms, a representative Python\tm library for writing quantum programs.
For mutation analysis of quantum programs, we use QMutpy~\cite{fortunato2022tqe}\footnote{\url{https://github.com/danielfobooss/mutpy}}, an extension of Mutpy, a mutation analysis tool for Python\tms.
The main contributions of this study are as follows:
\begin{itemize}
    \item We demonstrate that real-world faults pose a greater challenge for MBFL in quantum programs than artificial faults in terms of EXAM score\revision{, which represents the percentage of the code examined before locating the faulty statement}.
    \item We show that quantum mutation operations are effective for MBFL in quantum programs.
    \item We demonstrate the effectiveness of MBFL by comparing it with \textit{spectrum-based fault localization (SBFL)}, a method that relies on execution path information.
\end{itemize}


%% file: section/background.tex
\section{Related Work} \label{sec:related}

\noindent \textbf{Fault Localization for Classical Programs.}
Pearson et al.~\cite{pearson2017icse} compared the performance of SBFL (using five different formulas, \eg, Ochiai) and MBFL (MUSE and Metallaxis) on 2,995 artificially injected faults and 310 real-world faults in Java\tm programs.
Their results showed that while MBFL exhibited high performance for artificial faults (injected through mutation operations), it was less effective than SBFL in detecting real-world faults.
For quantum programs, no study has compared fault localization techniques from the perspectives of artificial injected faults vs. real-world faults, or SBFL vs. MBFL.
Consequently, it remains unclear which method is more effective for which types of faults.

\noindent \textbf{Fault Localization for Quantum Programs.}
Sato and Katsube~\cite{sato2024icse} identified four characteristics specific to quantum program testing (e.g., the cost of testing a specific part of a quantum program depends on its location) and proposed a fault localization method that accounts for these characteristics.
Their approach constructs a cost-based binary search tree from a quantum program and narrows down the testing scope using this tree to identify faulty locations.
This tree is built by treating the quantum program as a sequence of quantum gates.
However, real-world quantum programs, such as those in Bugs4Q~\cite{zhao2023jss}, often incorporate classical instructions (\eg, variable declarations and control structures) in addition to quantum instructions (\eg, quantum gate declarations and measurements).
As a result, representing a quantum program purely as a sequence of quantum gates may not always be feasible.
In contrast, mutation analysis can be applied to quantum programs containing both classical and quantum instructions.
Thus, we consider MBFL to have broader applicability.

\noindent \textbf{Mutation Analysis for Quantum Programs.}
Fortunato et al. \cite{fortunato2022tqe} developed QMutpy, a mutation analysis tool for quantum programs written in Qiskit\tms.
It extends Mutpy, an existing mutation tool for Python\tm programs.
They introduced five types of \textit{quantum mutation operations}: addition, deletion, and replacement of quantum gates, as well as addition and deletion of qubit measurements.
Since QMutpy is a fork of Mutpy, it also retains support for classical mutation operations (\eg, modifying arithmetic operators such as \texttt{+} to \texttt{-}).
Their experiments on 24 quantum programs demonstrated that quantum mutation operations resulted in higher mutation scores.
Another mutation analysis tool for quantum programs is Muskit~\cite{mendiluze2021ase}.
We use QMutpy because it provides more quantum mutation operations than Muskit and allows the combination of quantum and classical mutations.
Other studies have also conducted mutation analysis on quantum programs~\cite{ali2021icst,mendiluze2021ase, wang2022gecco}.
Their main goal is to expose misbehavior effectively and efficiently by mutation testing.
Our study differs from them in that it focuses on fault localization, aiming to identify the causes of bugs.



%% file: section/study_design.tex
\input{image/flow.tex}

\section{Study Design} \label{sec:method}
The goal of this study is to apply MBFL to quantum programs and evaluate its effectiveness.
We select MBFL as the target technique for two reasons: (1) Mutation testing for quantum programs has been actively studied and MBFL is considered a promising approach for quantum programs. (2) MBFL has a broad scope of applicability, as it can be applied to quantum programs that include both classical and quantum instructions.
Figure \ref{fig:flow} shows our study design.

\subsection{Research Questions}
\newcommand{\RQone}{\textit{\textbf{(Artificial vs. real-world faults)} How does the performance of MBFL differ between artificial and real-world faults?}}
\newcommand{\RQtwo}{\textit{\textbf{(Quantum vs. classical mutation operations)} Which is more effective, classical or quantum mutation operations?}}

\begin{enumerate}[label=\textit{RQ\arabic*}, leftmargin=*, align=left]
    \item \RQone
    \item \RQtwo
\end{enumerate}
RQ1 evaluates MBFL for quantum programs in terms of the type of faults (\ie, artificial vs. real-world).
It is similar to those in previous studies~\cite{pearson2017icse}, which evaluate the fault localization methods for Java\tm programs.
On the other hand, RQ2 is specific to quantum programs.
This RQ helps identify which types of mutation operations we should prioritize in our efforts.

\subsection{Bug Benchmarks}
We use both real-world and artificial bug benchmarks.
For the real-world benchmark, we use \textit{Bugs4Q}, proposed by Zhao et al.~\cite{zhao2023jss}.
For the artificial benchmark, we create \textit{BugsAqua}, which consists of bugs injected into Qiskit-Aqua\footnote{\url{https://github.com/qiskit-community/qiskit-aqua}} programs.

\subsubsection{Real-world Benchmark (Bugs4Q)}
Bugs4Q~\cite{zhao2023jss} is a benchmark consisting of 42 buggy programs written in Qiskit\textregistered.
These programs were collected from three popular platforms: GitHub\tms, Stack Overflow\tms, and Stack Exchange\tms.
For each buggy program, Bugs4Q provides the corresponding fixed version and test code to reproduce the bug. 
While the buggy and fixed programs were sourced from these platforms, the test code was manually written by the authors of the Bugs4Q paper.
Each test code consists of a single test case.
We use Bugs4Q because, to the best of our knowledge, it is the only benchmark that satisfies the following three criteria:
\begin{enumerate}
    \item The buggy programs are written by developers, meaning the benchmark contains real-world bugs.
    \item It includes both buggy and fixed versions of each program.
    \item It provides test code to reproduce the bugs.
\end{enumerate}
The second and third criteria are essential for evaluating fault localization results.
The test code is necessary to capture behavioral differences between the original program and its mutants, while the fixed program serves as a ground truth for faulty statements.

To reproduce the bugs in Bugs4Q, we cloned its replication package of Bugs4Q\footnote{\url{https://github.com/Z-928/Bugs4Q-Framework}} and executed the buggy and fixed programs.
For each sample in Bugs4Q, the test is expected to fail for the buggy version and pass for the fixed version.
However, 19 out of 42 bugs could not be reproduced under our experimental setup.
A possible reason is that the Bugs4Q paper and its replication package do not specify the versions of Python\tm and related libraries, such as Qiskit\tms, which may differ from our execution environment.
As a result, we exclude these non-reproducible cases and use 23 samples (i.e., 23 buggy/fixed programs and their test code) in our study.

\subsubsection{Artificial Benchmark (BugsAqua)}
\revision{We create BugsAqua, an artificial bug benchmark, by injecting faults into Qiskit-Aqua programs.}
The motivation for using an artificial benchmark is twofold.
First, since the number of the real-world bug benchmark is limited, incorporating an artificial bug benchmark allows us to evaluate the effectiveness of MBFL in a broader range of scenarios.
Second, artificial and real-world bug benchmarks may have different characteristics, which could help highlight key factors essential for addressing real-world faults.
We select Qiskit-Aqua as the target for fault injection because it has been used as an experimental target in the proposal paper for QMutPy~\cite{fortunato2022tqe}, a mutation tool for quantum programs. 
Qiskit-Aqua contains quantum programs that implement typical quantum algorithms and includes extensive test cases written by developers of Qiskit\tms.

Following their reproduction scripts\footnote{\url{https://github.com/jose/qmutpy-experiments}}, we selected the same 24 programs from the Qiskit-Aqua repository as their experimental targets. 
Each of these programs has a corresponding test code.
On average, each test code contains 36.2 test cases, with a minimum of 1 and a maximum of 593 test cases, which is significantly more than those in Bugs4Q.
We then apply QMutPy to these 24 programs, generating 2,361 mutants.
Among them, 594 mutants have at least one failing test case, indicating that the faults are successfully injected.
In this study, we exclude mutants whose execution time exceeds one hour.
This one hour threshold corresponds to a level that satisfies approximately 10\% of practitioners~\cite{kochhar2016issta}.
Although this level is relatively low, we set this timeout because a key focus of our study is to investigate the applicability and limitations of MBFL for quantum programs.
Programs with an execution time exceeding one hour are deemed too time-consuming\footnote{There were no cases in Bugs4Q where the execution time exceeded one hour.} for MBFL since it requires executing multiple mutants.
In fact, the estimated time for executing all the mutants can reach hundreds of hours in such cases.
As a result, we use the remaining 305 mutants as an artificial bug benchmark.
We refer to this benchmark as BugsAqua.
BugsAqua meets the same two criteria as Bugs4Q: (1) A fixed version of the buggy program is available (by comparing it with the original code), and (2) The test cases are provided (written by Qiskit\tm developers).

\subsection{Mutating Buggy Programs}
The first step in MBFL is to apply mutation operations to the buggy programs.
We use QMutPy to generate mutants.
In this study, we generate only first-order mutants, where a mutation operation is applied to a single statement in the program at a time.
QMutPy supports 20 classical and 5 quantum mutation operations.
We exclude two quantum mutation operations, quantum gate insertion and measurement insertion, because they do not work correctly for the programs in Bugs4Q\footnote{This issue arises because identifying the insertion positions for quantum gates and measurements using the abstract syntax tree does not work properly.}.
As a result, we apply a total of 23 mutation operations to the buggy programs.
QMutpy generated a total of 802 mutants for the 23 buggy programs in Bugs4Q and 34,627 mutants for the 305 buggy programs in BugsAqua.

The second step is to execute the test cases for the mutants.
We use the test results for both the buggy program and its mutants to localize faulty statements in the buggy program.

\subsection{Suspiciousness Scores}

MBFL calculates the \textit{suspiciousness score} for each statement based on the test results of the original program and its corresponding mutants.
The suspiciousness score indicates the likelihood that the statement contains a fault.
We use a simplified version of the formula proposed as MUSE~\cite{moon2014icst}, a representative MBFL method:
\begin{equation}
    S(s) = \frac{1}{| mut(s) |} \sum_{m \in mut(s)} \frac{| f_P(s) \cap p_m |}{| f_P(s) |} . \label{eq:susp}
\end{equation}
Here, $mut(s)$ represents the set of mutants generated by applying all mutation operations to statement 
$s$.
$f_P(s)$ represents the set of failed test cases when statement $s$ is executed in program $P$.
Similarly, $p_m$ represents the set of passed test cases in $m$, where $m$ is a mutant of statement $s$.
The suspiciousness score for $s$ increases when mutating $s$ frequently changes failing test cases to passing.



\subsection{Evaluation Metrics for Fault Localization}
To evaluate the output of fault localization method (\ie, a ranked list of statements by suspiciousness scores in descending order), we use \textit{EXAM} score, commonly used in existing studies~\cite{ju2014jss, pearson2017icse}.
This score is defined as ``the percentage of statements in a program that have to be examined until the first faulty statement is reached''~\cite{wong2016tse}:
\begin{equation}
    EXAM = \frac{\text{Rank of the faulty statement}}{\text{Total number of statements in a program}} \times 100\% . \label{eq:exam}
\end{equation}
The lower this percentage is, the more effective the method is, since it allows developers to find faulty statements with less effort.

In our experiments, the ground truth for the faulty statement is obtained from the fixed version of a buggy program in Bugs4Q and from the original version of a fault-injected program in BugsAqua, respectively.
In BugsAqua, each buggy program contains only one faulty statement, whereas Bugs4Q may have multiple faulty statements, as it is a real-world bug benchmark.
In the latter case, we use the rank of the highest-ranked faulty statement among the multiple faulty statements, as the EXAM score is defined based on the position of the ``first'' faulty statement.

Besides, there may be multiple statements with the same suspiciousness score as the faulty statement.
In such cases, we adjust the calculation of the rank, which serves as the numerator in Equation \eqref{eq:exam}.
As suggested in a survey paper on fault localization~\cite{wong2016tse}, we report both the \textit{best-case} and \textit{worst-case} scenarios.
In the best-case scenario, the faulty statement is assumed to be the first one found when checking all statements with the same suspiciousness score sequentially.
Similarly, in the worst-case scenario, the faulty statement is assumed to be the last one found.
If the rank of the faulty statement is $r$ and the number of statements sharing the same suspiciousness score is $n$, then the rank used in the best-case and worst-case scenarios is $r$ and $r+n-1$, respectively. 

\subsection{Experimental Environment}
All our experiments are performed on a classical computer.
Although programs written in Qiskit\tm can be executed on actual quantum computers, they can also be executed on classical computers as the quantum computer simulators.
We used this simulator functionality because of the hurdles posed by the availability and noise of real quantum computers.
The Python\tm version used is 3.9.0.
The versions of related to Qiskit\tm are as follows: \textit{qiskit-aer: 0.10.0}, \textit{qiskit-aqua: 0.9.5}, \textit{qiskit-ignis: 0.7.1}, \textit{qiskit-terra: 0.20.0}.

%% file: image/flow.tex
\begin{figure}[t]
    \begin{center}
      \includegraphics[width=0.9\linewidth]{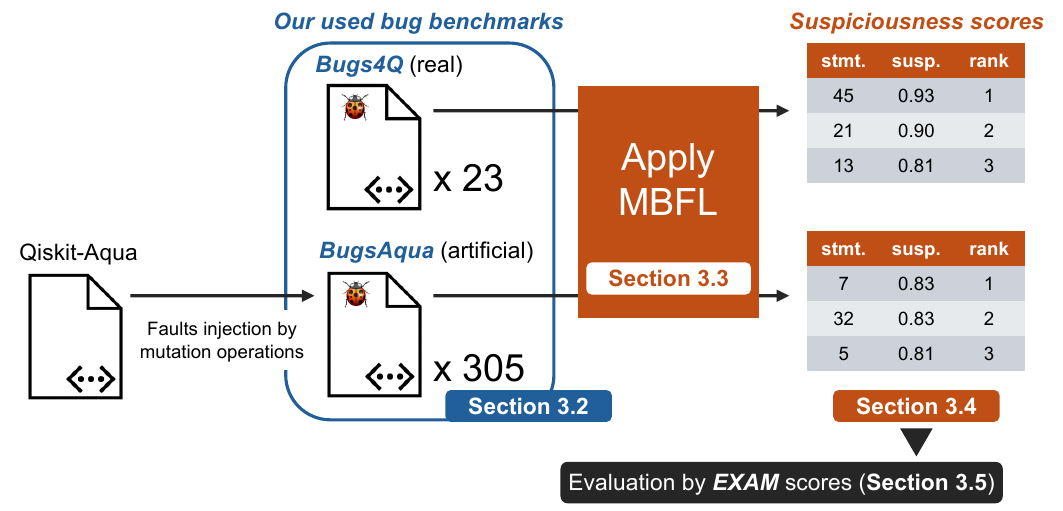} 
      \caption{Our study design.}
      \label{fig:flow}
    \end{center}
  \end{figure}

%% file: section/results.tex
\section{Results} \label{sec:results}

\input{section/rqs/rq1.tex}
\input{section/rqs/rq2.tex}

%% file: section/rqs/rq1.tex
\subsection{RQ1: Artificial vs. Real-world Faults}
\input{image/rq1_exam.tex}

\fig{fig:rq1_exam} presents the empirical cumulative distribution function of EXAM scores for Bugs4Q and BugsAqua.
This figure indicates that real-world faults are more challenging for MBFL than artificial faults because the EXAM scores are higher for Bugs4Q than for BugsAqua.
The median EXAM score in the best-case scenario is 8.7\% for Bugs4Q and 0.9\% for BugsAqua.
In the worst-case scenario, the median EXAM score is 19.4\% for Bugs4Q and 1.2\% for BugsAqua.
This finding is consistent with the results for Java\tm program conducted by Pearson~\et~\cite{pearson2017icse}.
One possible explanation for this result is the presence of ``reversible'' mutants, also reported in the study of Pearson~\et~\cite{pearson2017icse}.
In our study, for example, an artificial fault injected by replacing a quantum gate can be fixed by applying a mutation operation that replaces it back with the original quantum gate.
In contrast, real-world faults were less likely to be ``reversible'' through such simple mutation operations.

The range of EXAM scores for Bugs4Q is wider than that for BugsAqua (\ie, filled areas in \fig{fig:rq1_exam}), indicating greater variability in MBFL performance for Bugs4Q.
In Bugs4Q, the worst-case EXAM score is 100\% in approximately 40\% of cases.
This suggests that MBFL may fail in many cases.
The reason for this can be that faults in Bugs4Q are too complex to detect with simple mutation operations \revision{as discussed in \sec{sec:dis_challenge}}.
\revision{
Furthermore, test case quality may impact MBFL performance. 
Bugs4Q provides only one test case per bug, written by its authors rather than the original developers. 
These test cases may lack robustness to capture behavioral differences caused by mutants, potentially leading to poor results.
}

\summarybox{Answer to RQ1}{
Real-world faults are more challenging for MBFL than artificial faults.
In the worst-case scenario, the median EXAM score is 19.4\% for Bugs4Q and 1.2\% for BugsAqua.
Additionaly, the performance of MBFL for Bugs4Q is more unstable than that for BugsAqua.
}

%% file: image/rq1_exam.tex
\begin{figure}[t]
    \begin{center}
      \includegraphics[width=0.8\linewidth]{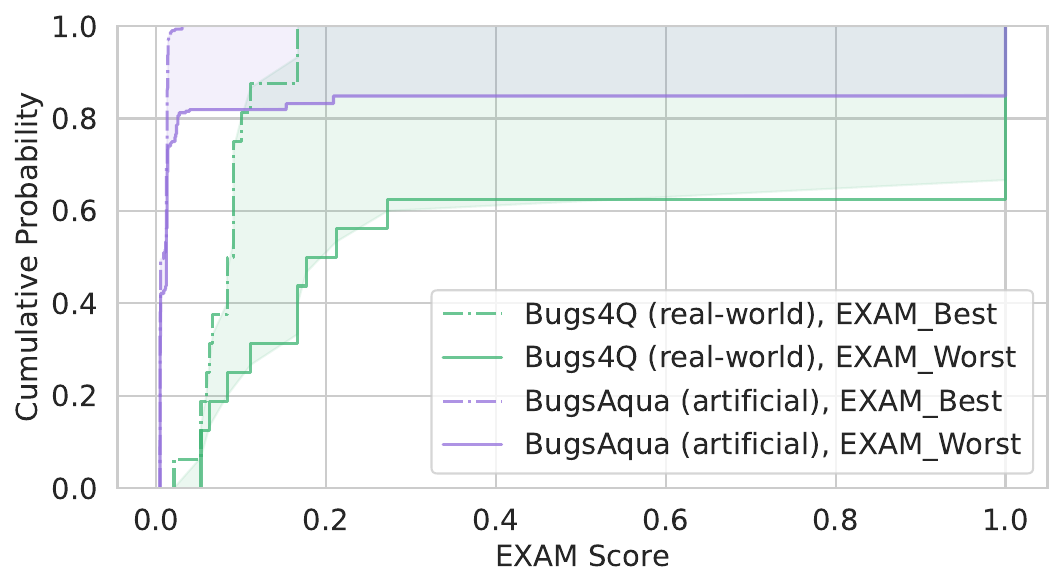}
      \caption{Empirical cumulative distribution function of EXAM scores for real-world faults (Bugs4Q) and artificial faults (BugsAqua).
      The filled areas indicate the range between the best-case and worst-case EXAM scores.}
      \label{fig:rq1_exam}
    \end{center}
  \end{figure}

%% file: section/rqs/rq2.tex
\subsection{RQ2: Quantum vs. Classical Mutation Operations}

\input{table/rq2_summary.tex}
As shown in Equation \eqref{eq:susp}, the success of MBFL depends on the presence of mutants that change test results (\ie, \textit{behavior-changing mutants}).
Therefore, we examine the number of behavior-changing mutants for each category of mutation operation (\ie, quantum or classical).
This analysis helps us understand the effectiveness of quantum and classical mutation operations for MBFL.

\tab{tab:rq2} shows the numbers of behavior-changing mutants in Bugs4Q and BugsAqua.
For Bugs4Q, most of the behavior-changing mutants are quantum mutants (31 out of 33 mutants, 93.9\%).
It indicates that for Bugs4Q, the success of MBFL heavily depends on the use of quantum mutation operations.
On the other hand, for BugsAqua, a larger proportion of behavior-changing mutants are classical mutants (302 out of 502 mutants, 60.2\%).
This is because faults in BugsAqua, particularly those injected by classical mutation operations, are often reversible using classical mutation operations.

Per mutation operation, quantum mutation operations generate more behavior-changing mutants than classical mutation operations.
For Bugs4Q, quantum mutation operations generate 10.3 behavior-changing mutants per operation, while classical mutation operations generate only 0.1 per operation.
A similar trend is observed for BugsAqua (see line for Avg. per Op. in \tab{tab:rq2}).

\summarybox{Answer to RQ2}{
The results suggest that quantum mutation operations are more effective than classical mutation operations in changing test results.
For Bugs4Q, 93.9\% of the behavior-changing mutants are quantum mutants.
}

%% file: table/rq2_summary.tex

\begin{table}[t]
    \begin{center}
    \caption{Numbers of behavior-changing mutants (shown in the column \#Mutants (B.-C.)). \revision{The column Avg. per Op. is calculated as \#Mutants (B.-C.) divided by \#Ops.}}
    \scalebox{0.8}{
    \begin{tabular}{lllll}
        \toprule
        Benchmark & Mutation Type & \#Ops. & \#Mutants (B.-C.) & Avg. per Op.\\
        \midrule
        \multirow{2}{*}{Bugs4Q}  
            & Quantum  & 3  & 31 (93.9\%) & 10.3 \\
            & Classical & 20  & 2 (6.1\%) & 0.1\\
        \midrule
        \multirow{2}{*}{BugsAqua}  
            & Quantum  & 3  & 200 (39.8\%) & 66.7\\
            & Classical & 20  & 302 (60.2\%) & 15.1\\
        \bottomrule
    \end{tabular}
    }
    \label{tab:rq2}
    \end{center}
\end{table}

%% file: section/discussion.tex
\section{Discussion} \label{sec:discussion}

\subsection{Why are real-world faults in quantum programs challenging for MBFL?} \label{sec:dis_challenge}
In this section, we compare cases where MBFL succeeded and failed in Bugs4Q.
\revision{This comparison allows us to discuss the boundaries of faults that existing quantum mutation operations can and cannot handle.
Furthermore, we explore future directions for enhancing MBFL in quantum programs.}

\revision{Listing~\ref{lst:bug1}} is an example where MBFL assigned the highest suspiciousness score to the faulty statement.
\begin{lstlisting}[
    language=Python,
    caption={A code fragment from id=1 in Bugs4Q.},
    label={lst:bug1},
]
qc = QuantumCircuit(3)
qc.cx(0, 1, label='Label', ctrl_state=0)
qc.ccx(0, 1, 2, label='Label', ctrl_state=1) # This line causes an error.
\end{lstlisting}
This fault occurred because the developer incorrectly specified the arguments for the \texttt{ccx} gate.
Such a fault can be detected using a mutation operation that deletes the quantum gate.

\revision{Listing~\ref{lst:bug39}} is an example where MBFL failed to assign a high suspiciousness score to the faulty statement.
\begin{lstlisting}[
    language=Python,
    caption={A code fragment from id=39 in Bugs4Q.},
    label={lst:bug39},
]
qc = QuantumCircuit(4, 4)
# for i in range(4): # These lines are 
#     qc.h(i)        # added in the fixed ver.
qc.cx(3, 1)
qc.cx(1, 0)
qc.cx(0, 1)
qc.ccx(3, 2, 1)
qc.cx(1, 2)
qc.cx(3, 2)
qc.measure(0, 0)
\end{lstlisting}
This fault occurred because the developer forgot to initialize all four qubits with the \texttt{h} gate.
The \texttt{h} gate transforms the quantum state \(|0\rangle\) into \(\frac{1}{\sqrt{2}}(|0\rangle + |1\rangle)\), creating a superposition of 0 and 1.
This type of initialization is commonly used in many well-known quantum algorithms~\cite{nielsen2010quantum}.
MBFL with existing mutation operations cannot handle this type of fault \revision{(\ie, \textit{pattern-related faults})}.

Two possible approaches to address this issue are (1) using \textit{high-order mutants (HOM)}, which apply mutation operations at multiple locations simultaneously, and (2) introducing this type of initialization as a new mutation operation because it is a commonly used pattern.
We consider (2) to be a promising direction because HOM introduce efficiency challenges, such as a significant increase in the number of mutants~\cite{jia2009ist}.
Collaborating with experts for quantum computing or application domains (e.g., chemistry) could help identify such types of frequent operation patterns.
Alternatively, mining version control histories could reveal common operation patterns that developers make in practice.

\summarybox{Future directions}{
Enriching quantum mutation operations would be a next step toward effectively detecting real-world faults in quantum programs.
For instance, identifying common operation patterns in quantum programs (\eg, initialization) and introducing them as new mutation operations could enhance the effectiveness of MBFL.
}

\input{table/sbfl_vs_mbfl.tex}

\subsection{Comparison between MBFL and SBFL}
In this section, we compare the performance of MBFL and SBFL for the quantum programs.
Unlike MBFL, SBFL does not rely on mutants; it utilizes differences in execution paths.
This comparison allows us to investigate the benefits of MBFL gained from utilizing mutants in addition to execution paths.
We use two representative formulas for SBFL, \ie, Ochiai~\cite{abreu2007ochiai} and Tarantula~\cite{jones2005ase}.
A key difference from MBFL is that they do not use \( mut(s) \) like Equation \eqref{eq:susp}.

For Bugs4Q, SBFL cannot be applied because there is only one failing test case per bug.
Since SBFL calculates suspiciousness scores based on differences in execution paths, it cannot be applied when there is only a single test case.
MBFL can still be applied in such cases, showing its broader applicability compared to SBFL.

For BugsAqua, we compare the EXAM scores of MBFL and SBFL for each of the 305 buggy programs.
Table~\ref{tab:mbfl_vs_sbfl} shows the comparison results of MBFL vs. SBFL{\scriptsize{Ochiai}} and MBFL vs. SBFL{\scriptsize{Tarantula}}.
For each comparison, we conducted a Wilcoxon signed-rank test~\cite{conover1999stat} on both the best-case and worst-case EXAM scores.
Since we hypothesize that MBFL outperforms SBFL, we applied a one-sided test to determine whether MBFL yields lower EXAM scores.
Additionally, we assessed the effect size using Cliff’s $\delta$~\cite{cliff1993dominance}.
In this case, a negative \( \delta \) indicates the extent to which the EXAM score for MBFL is lower compared to that for SBFL.

From Table~\ref{tab:mbfl_vs_sbfl}, we observe that in the best-case scenario, the difference between MBFL and SBFL is negligible.
In contrast, in the worst-case scenario, MBFL tends to achieve a significantly lower EXAM score compared to SBFL.
Specifically, both comparisons exhibit statistical significance with $p < 0.01$ and a large effect size according to a guideline by Kitchenham~\et~\cite{kitchenham2017emse}.
According to the definition of the worst-case EXAM score, these results suggest that SBFL tends to assign the same suspiciousness score to a larger number of statements, including the faulty statement.
It indicates the unstable performance of SBFL compared to MBFL.

\summarybox{Summary}{
For Bugs4Q, SBFL cannot be applied because there is only one failing test case per bug.
For BugsAqua, the performance of MBFL is not significantly different from SBFL in the best-case scenario.
However, in the worst-case scenario, MBFL achieves a lower EXAM score than SBFL, indicating the performance instability of SBFL.
}

%% file: table/sbfl_vs_mbfl.tex
\begin{table}[t]
    \begin{center}
    \caption{Comparison of EXAM scores between MBFL and SBFL methods.
    In the ``Sig. level'' column,  ``**'' and ``-'' indicate $p < 0.01$ and $p > 0.05$, respectively.}
    \scalebox{0.75}{ 
    \begin{tabular}{lccc}
        \toprule
        Comparison & EXAM & Cliff's $\delta$ & Sig. level \\
        \midrule
        \multirow{2}{*}{MBFL vs. SBFL\scriptsize{Ochiai}}  
            & best & -0.0106 & - \\
            & worst & -0.7491 & ** \\
        \midrule
        \multirow{2}{*}{MBFL vs. SBFL\scriptsize{Trantula}}  
            & best & 0.0001 & - \\
            & worst & -0.7491 & ** \\
        \bottomrule
    \end{tabular}
    }
    \label{tab:mbfl_vs_sbfl}
    \end{center}
\end{table}

%% file: section/validity.tex
\section{Threats to Validity} \label{sec:threats}

\noindent \textbf{Construct validity:}
Excluding programs in our study design (\eg, due to reproducibility issues or timeouts) may have introduced bias in the benchmarks.
Furthermore, evaluating the performance of fault localization methods using only the EXAM score may not be sufficient.
While we aimed for a comprehensive evaluation by considering both the best-case and worst-case scenarios, using additional evaluation metrics could provide more robust results.

\noindent \textbf{Internal validity:} 
We made minor modifications to the source code of QMutPy to make it compatible with MBFL.
For example, QMutpy exits with an error if all tests for the program under test fail, as it is originally designed for mutation testing. 
In the case of MBFL, we would like the mutation operation to be applied even in such cases. 
The minor modifications were necessary to achieve this.
While these changes were minimal, they may have inadvertently affected other parts of the tool.

\noindent \textbf{External validity:}
Since we ran quantum programs on a simulator by Qiskit\tms, our findings may not be generalizable to real quantum computers.
Moreover, it is unclear whether our findings would generalize to quantum programs written in frameworks other than Qiskit\tm or in programming languages other than Python\tms.





%% file: section/conclusion.tex
\section{Conclusion} \label{sec:conclusion}
This study evaluates the effectiveness of MBFL for quantum programs written in Qiskit\tm using both real-world and artificial bug benchmarks.
The results of RQ1 show that the EXAM score for BugsAqua is lower than that for Bugs4Q in both best-case and worst-case scenarios.
This findings suggest that MBFL is more effective for artificial faults than for real-world faults.
The results of RQ2 indicate that quantum mutation operations are more effective than classical mutation operations in changing test results.
For Bugs4Q, 93.9\% of the mutants that change test results are generated by quantum mutation operations.
We also discuss the future directions by diving deeper into the successes and failures of MBFL.
One possible direction is to enrich quantum mutation operations and improve the effectiveness of MBFL for quantum programs.




%% file: main.bbl

\begin{thebibliography}{22}


\ifx \showCODEN    \undefined \def \showCODEN     #1{\unskip}     \fi
\ifx \showISBNx    \undefined \def \showISBNx     #1{\unskip}     \fi
\ifx \showISBNxiii \undefined \def \showISBNxiii  #1{\unskip}     \fi
\ifx \showISSN     \undefined \def \showISSN      #1{\unskip}     \fi
\ifx \showLCCN     \undefined \def \showLCCN      #1{\unskip}     \fi
\ifx \shownote     \undefined \def \shownote      #1{#1}          \fi
\ifx \showarticletitle \undefined \def \showarticletitle #1{#1}   \fi
\ifx \showURL      \undefined \def \showURL       {\relax}        \fi
\providecommand\bibfield[2]{#2}
\providecommand\bibinfo[2]{#2}
\providecommand\natexlab[1]{#1}
\providecommand\showeprint[2][]{arXiv:#2}

\bibitem[Abreu et~al\mbox{.}(2007)]%
        {abreu2007ochiai}
\bibfield{author}{\bibinfo{person}{Rui Abreu}, \bibinfo{person}{Peter Zoeteweij}, {and} \bibinfo{person}{Arjan~JC Van~Gemund}.} \bibinfo{year}{2007}\natexlab{}.
\newblock \showarticletitle{On the accuracy of spectrum-based fault localization}. In \bibinfo{booktitle}{\emph{Testing: Academic and industrial conference practice and research techniques-MUTATION (TAICPART-MUTATION 2007)}}. IEEE, \bibinfo{pages}{89--98}.
\newblock


\bibitem[Ali et~al\mbox{.}(2021)]%
        {ali2021icst}
\bibfield{author}{\bibinfo{person}{Shaukat Ali}, \bibinfo{person}{Paolo Arcaini}, \bibinfo{person}{Xinyi Wang}, {and} \bibinfo{person}{Tao Yue}.} \bibinfo{year}{2021}\natexlab{}.
\newblock \showarticletitle{Assessing the effectiveness of input and output coverage criteria for testing quantum programs}. In \bibinfo{booktitle}{\emph{Proceedings of the 14th IEEE Conference on Software Testing, Verification and Validation}}. \bibinfo{pages}{13--23}.
\newblock


\bibitem[Chen et~al\mbox{.}(2023)]%
        {chen2023icse}
\bibfield{author}{\bibinfo{person}{Qihong Chen}, \bibinfo{person}{R{\'u}ben C{\^a}mara}, \bibinfo{person}{Jos{\'e} Campos}, \bibinfo{person}{Andr{\'e} Souto}, {and} \bibinfo{person}{Iftekhar Ahmed}.} \bibinfo{year}{2023}\natexlab{}.
\newblock \showarticletitle{The smelly eight: An empirical study on the prevalence of code smells in quantum computing}. In \bibinfo{booktitle}{\emph{Proceedings of the IEEE/ACM 45th International Conference on Software Engineering}}. \bibinfo{pages}{358--370}.
\newblock


\bibitem[Cliff(1993)]%
        {cliff1993dominance}
\bibfield{author}{\bibinfo{person}{Norman Cliff}.} \bibinfo{year}{1993}\natexlab{}.
\newblock \showarticletitle{Dominance statistics: Ordinal analyses to answer ordinal questions.}
\newblock \bibinfo{journal}{\emph{Psychological bulletin}} \bibinfo{volume}{114}, \bibinfo{number}{3} (\bibinfo{year}{1993}), \bibinfo{pages}{494}.
\newblock


\bibitem[Conover(1999)]%
        {conover1999stat}
\bibfield{author}{\bibinfo{person}{William~Jay Conover}.} \bibinfo{year}{1999}\natexlab{}.
\newblock \bibinfo{booktitle}{\emph{Practical nonparametric statistics}}.
\newblock \bibinfo{publisher}{john wiley \& sons}.
\newblock


\bibitem[Fortunato et~al\mbox{.}(2022)]%
        {fortunato2022tqe}
\bibfield{author}{\bibinfo{person}{Daniel Fortunato}, \bibinfo{person}{Jose Campos}, {and} \bibinfo{person}{Rui Abreu}.} \bibinfo{year}{2022}\natexlab{}.
\newblock \showarticletitle{Mutation testing of quantum programs: A case study with Qiskit}.
\newblock \bibinfo{journal}{\emph{IEEE Transactions on Quantum Engineering}}  \bibinfo{volume}{3} (\bibinfo{year}{2022}), \bibinfo{pages}{1--17}.
\newblock


\bibitem[Ishimoto et~al\mbox{.}(2024)]%
        {ishimoto2024apsec}
\bibfield{author}{\bibinfo{person}{Yuta Ishimoto}, \bibinfo{person}{Yuto Nakamura}, \bibinfo{person}{Ryota Katsube}, \bibinfo{person}{Naoto Sato}, \bibinfo{person}{Hideto Ogawa}, \bibinfo{person}{Masanari Kondo}, \bibinfo{person}{Yasutaka Kamei}, {and} \bibinfo{person}{Naoyasu Ubayashi}.} \bibinfo{year}{2024}\natexlab{}.
\newblock \showarticletitle{An Empirical Study on Self-Admitted Technical Debt in Quantum Software}. In \bibinfo{booktitle}{\emph{Proceedings of the 31st Asia-Pacific Software Engineering Conference (APSEC)}}. \bibinfo{pages}{41--50}.
\newblock


\bibitem[Jia and Harman(2009)]%
        {jia2009ist}
\bibfield{author}{\bibinfo{person}{Yue Jia} {and} \bibinfo{person}{Mark Harman}.} \bibinfo{year}{2009}\natexlab{}.
\newblock \showarticletitle{Higher order mutation testing}.
\newblock \bibinfo{journal}{\emph{Information and Software Technology}} \bibinfo{volume}{51}, \bibinfo{number}{10} (\bibinfo{year}{2009}), \bibinfo{pages}{1379--1393}.
\newblock


\bibitem[Jones and Harrold(2005)]%
        {jones2005ase}
\bibfield{author}{\bibinfo{person}{James~A Jones} {and} \bibinfo{person}{Mary~Jean Harrold}.} \bibinfo{year}{2005}\natexlab{}.
\newblock \showarticletitle{Empirical evaluation of the tarantula automatic fault-localization technique}. In \bibinfo{booktitle}{\emph{Proceedings of the 20th IEEE/ACM international Conference on Automated software engineering}}. \bibinfo{pages}{273--282}.
\newblock


\bibitem[Ju et~al\mbox{.}(2014)]%
        {ju2014jss}
\bibfield{author}{\bibinfo{person}{Xiaolin Ju}, \bibinfo{person}{Shujuan Jiang}, \bibinfo{person}{Xiang Chen}, \bibinfo{person}{Xingya Wang}, \bibinfo{person}{Yanmei Zhang}, {and} \bibinfo{person}{Heling Cao}.} \bibinfo{year}{2014}\natexlab{}.
\newblock \showarticletitle{HSFal: Effective fault localization using hybrid spectrum of full slices and execution slices}.
\newblock \bibinfo{journal}{\emph{Journal of Systems and Software}}  \bibinfo{volume}{90} (\bibinfo{year}{2014}), \bibinfo{pages}{3--17}.
\newblock


\bibitem[Kitchenham et~al\mbox{.}(2017)]%
        {kitchenham2017emse}
\bibfield{author}{\bibinfo{person}{Barbara Kitchenham}, \bibinfo{person}{Lech Madeyski}, \bibinfo{person}{David Budgen}, \bibinfo{person}{Jacky Keung}, \bibinfo{person}{Pearl Brereton}, \bibinfo{person}{Stuart Charters}, \bibinfo{person}{Shirley Gibbs}, {and} \bibinfo{person}{Amnart Pohthong}.} \bibinfo{year}{2017}\natexlab{}.
\newblock \showarticletitle{Robust statistical methods for empirical software engineering}.
\newblock \bibinfo{journal}{\emph{Empirical Software Engineering}}  \bibinfo{volume}{22} (\bibinfo{year}{2017}), \bibinfo{pages}{579--630}.
\newblock


\bibitem[Kochhar et~al\mbox{.}(2016)]%
        {kochhar2016issta}
\bibfield{author}{\bibinfo{person}{Pavneet~Singh Kochhar}, \bibinfo{person}{Xin Xia}, \bibinfo{person}{David Lo}, {and} \bibinfo{person}{Shanping Li}.} \bibinfo{year}{2016}\natexlab{}.
\newblock \showarticletitle{Practitioners' expectations on automated fault localization}. In \bibinfo{booktitle}{\emph{Proceedings of the 25th international symposium on software testing and analysis}}. \bibinfo{pages}{165--176}.
\newblock


\bibitem[Mendiluze et~al\mbox{.}(2021)]%
        {mendiluze2021ase}
\bibfield{author}{\bibinfo{person}{E{\~n}aut Mendiluze}, \bibinfo{person}{Shaukat Ali}, \bibinfo{person}{Paolo Arcaini}, {and} \bibinfo{person}{Tao Yue}.} \bibinfo{year}{2021}\natexlab{}.
\newblock \showarticletitle{Muskit: A mutation analysis tool for quantum software testing}. In \bibinfo{booktitle}{\emph{Proceedings of the 36th IEEE/ACM International Conference on Automated Software Engineering}}. \bibinfo{pages}{1266--1270}.
\newblock


\bibitem[Moon et~al\mbox{.}(2014)]%
        {moon2014icst}
\bibfield{author}{\bibinfo{person}{Seokhyeon Moon}, \bibinfo{person}{Yunho Kim}, \bibinfo{person}{Moonzoo Kim}, {and} \bibinfo{person}{Shin Yoo}.} \bibinfo{year}{2014}\natexlab{}.
\newblock \showarticletitle{Ask the mutants: Mutating faulty programs for fault localization}. In \bibinfo{booktitle}{\emph{Proceedings of the IEEE Seventh International Conference on Software Testing, Verification and Validation}}. \bibinfo{pages}{153--162}.
\newblock


\bibitem[Nielsen and Chuang(2010)]%
        {nielsen2010quantum}
\bibfield{author}{\bibinfo{person}{Michael~A Nielsen} {and} \bibinfo{person}{Isaac~L Chuang}.} \bibinfo{year}{2010}\natexlab{}.
\newblock \bibinfo{booktitle}{\emph{Quantum computation and quantum information}}.
\newblock \bibinfo{publisher}{Cambridge university press}.
\newblock


\bibitem[Pearson et~al\mbox{.}(2017)]%
        {pearson2017icse}
\bibfield{author}{\bibinfo{person}{Spencer Pearson}, \bibinfo{person}{Jos{\'e} Campos}, \bibinfo{person}{Ren{\'e} Just}, \bibinfo{person}{Gordon Fraser}, \bibinfo{person}{Rui Abreu}, \bibinfo{person}{Michael~D Ernst}, \bibinfo{person}{Deric Pang}, {and} \bibinfo{person}{Benjamin Keller}.} \bibinfo{year}{2017}\natexlab{}.
\newblock \showarticletitle{Evaluating and improving fault localization}. In \bibinfo{booktitle}{\emph{Proceedings of the IEEE/ACM 39th International Conference on Software Engineering}}. \bibinfo{pages}{609--620}.
\newblock


\bibitem[Ramalho et~al\mbox{.}(2024)]%
        {ramalho2024arxiv}
\bibfield{author}{\bibinfo{person}{Neilson Carlos~Leite Ramalho}, \bibinfo{person}{Higor~Amario de Souza}, {and} \bibinfo{person}{Marcos~Lordello Chaim}.} \bibinfo{year}{2024}\natexlab{}.
\newblock \showarticletitle{Testing and Debugging Quantum Programs: The Road to 2030}.
\newblock \bibinfo{journal}{\emph{arXiv preprint arXiv:2405.09178}} (\bibinfo{year}{2024}).
\newblock


\bibitem[Sato and Katsube(2024)]%
        {sato2024icse}
\bibfield{author}{\bibinfo{person}{Naoto Sato} {and} \bibinfo{person}{Ryota Katsube}.} \bibinfo{year}{2024}\natexlab{}.
\newblock \showarticletitle{Locating Buggy Segments in Quantum Program Debugging}. In \bibinfo{booktitle}{\emph{Proceedings of the ACM/IEEE 44th International Conference on Software Engineering: New Ideas and Emerging Results}}. \bibinfo{pages}{26--31}.
\newblock


\bibitem[Shaydulin et~al\mbox{.}(2020)]%
        {shaydulin2020icsew}
\bibfield{author}{\bibinfo{person}{Ruslan Shaydulin}, \bibinfo{person}{Caleb Thomas}, {and} \bibinfo{person}{Paige Rodeghero}.} \bibinfo{year}{2020}\natexlab{}.
\newblock \showarticletitle{Making Quantum Computing Open: Lessons from Open Source Projects}. In \bibinfo{booktitle}{\emph{Proceedings of the IEEE/ACM 42nd International Conference on Software Engineering Workshops}}. \bibinfo{pages}{451–455}.
\newblock


\bibitem[Wang et~al\mbox{.}(2022)]%
        {wang2022gecco}
\bibfield{author}{\bibinfo{person}{Xinyi Wang}, \bibinfo{person}{Tongxuan Yu}, \bibinfo{person}{Paolo Arcaini}, \bibinfo{person}{Tao Yue}, {and} \bibinfo{person}{Shaukat Ali}.} \bibinfo{year}{2022}\natexlab{}.
\newblock \showarticletitle{Mutation-based test generation for quantum programs with multi-objective search}. In \bibinfo{booktitle}{\emph{Proceedings of the genetic and evolutionary computation conference}}. \bibinfo{pages}{1345--1353}.
\newblock


\bibitem[Wong et~al\mbox{.}(2016)]%
        {wong2016tse}
\bibfield{author}{\bibinfo{person}{W~Eric Wong}, \bibinfo{person}{Ruizhi Gao}, \bibinfo{person}{Yihao Li}, \bibinfo{person}{Rui Abreu}, {and} \bibinfo{person}{Franz Wotawa}.} \bibinfo{year}{2016}\natexlab{}.
\newblock \showarticletitle{A survey on software fault localization}.
\newblock \bibinfo{journal}{\emph{IEEE Transactions on Software Engineering}} \bibinfo{volume}{42}, \bibinfo{number}{8} (\bibinfo{year}{2016}), \bibinfo{pages}{707--740}.
\newblock


\bibitem[Zhao et~al\mbox{.}(2023)]%
        {zhao2023jss}
\bibfield{author}{\bibinfo{person}{Pengzhan Zhao}, \bibinfo{person}{Zhongtao Miao}, \bibinfo{person}{Shuhan Lan}, {and} \bibinfo{person}{Jianjun Zhao}.} \bibinfo{year}{2023}\natexlab{}.
\newblock \showarticletitle{Bugs4Q: A benchmark of existing bugs to enable controlled testing and debugging studies for quantum programs}.
\newblock \bibinfo{journal}{\emph{Journal of Systems and Software}}  \bibinfo{volume}{205} (\bibinfo{year}{2023}), \bibinfo{pages}{111805}.
\newblock


\end{thebibliography}
